# Adaptive Interaction Using the Adaptive Agent Oriented Software Architecture (AAOSA)

Babak Hodjat          Makoto Amamiya

Department of Intelligent Systems
Graduate School of Information Science and Electrical Engineering
Kyushu University
6-1 Kasugakoen, Kasuga-shi
Fukuoka 816, Japan
http://www_al.is.kyushu-u.ac.jp/~bobby/index.html
E-mail: Bobby@diana.is.kyushu-u.ac.jp

**Abstract**

User interfaces that adapt their characteristics to those of the user are referred to as adaptive interfaces. We propose Adaptive Agent Oriented Software Architecture (AAOSA) as a new way of designing adaptive interfaces. AAOSA is a new approach to software design based on an agent-oriented architecture. In this approach agents are considered adaptively communicating concurrent modules which are divided into a white box module responsible for the communications and learning, and a black box which is responsible for the independent specialized processes of the agent. A distributed learning policy that makes use of this architecture is used for purposes of system adaptability.

**Keywords**: Adaptive interfaces, Agent-oriented systems, Multi-Agent Software architectures, Distributed Learning,

## 1. Introduction

Most of the human-computer interfaces being used today are complicated and difficult to use. This is due mostly to the growing number of features the interface should provide easy access to.
Users usually have the following problems with current interface [1]:
- Prior to selecting an action. They have to consider if the application provides an appropriate action at all. This hints on a need for some sort of feedback from the application.
- It is hard to access the actions they already know about. This implies that the user should be able to freely express his or her needs without being bound to the limited conventions preset by the application.
- They have to imagine what would be an appropriate action to proceed with in order to perform a certain task of the application domain. The application, therefore, should be able to guide the users through the many options they may at any stage of the interaction.

The objective of this paper is to propose a user interface that will overcome the above difficulties. User interfaces that adapt their characteristics to those of the user are referred to as adaptive interfaces. These interactive software systems improve their ability to interact with a user based on partial experience with that user. The user's decisions give a ready source of training data to support learning. Every time the interface suggests some choice, the human either accepts that recommendation or rejects it, whether this feedback is explicit or simply reflected in the user's behavior. Before detailing our proposal any further, let us see what general features may be desirable in a user interface:

➢ Natural expression: The user should be able to express his or her intentions as freely and naturally as possible.
➢ Optimum interaction: Interaction should be limited to the following:
  - The user is in doubt as to what she/he can do next or how she/he can do it.
  - The system is in doubt as to what the user intends to do next.
  
  We must stress here that lack of interaction or feedback from the system is not at all desirable. By Optimum interaction we mean interaction where it is required, no more and no less.

- Adaptability: Adaptability could be about the changing context of interaction or application, but most importantly, the system should be able to adapt to the user's way of expressing her/his intentions. Two main issues that will have to be taken into account in this regard are generalization and contradiction recovery.
    - Generalization: An adaptable system in its simplest form will only learn the instance that it has been taught (implicitly or explicitly). Generalization occurs when the system uses what it has learned, to resolve problems it deems similar. The success and degree of generalization, therefore, depend directly on the precision of the similarity function and the threshold the system uses to distinguish between similar and dissimilar situations.
    - Contradiction: A system that generalizes may well over-generalize. The moment the system's reaction based on a generalization is in a manner the user does not anticipate, the system has run into a contradiction. The resolution of this contradiction is an integral part of the learning and adaptability process.
- Ease of change and upgrade: The application designer should easily be able to upgrade or change the system with minimum compromise to the adaptation the system has made to users. This change should be done at run-time (i.e., on the fly).

Various attempts have been made to reduce the navigation effort in menu hierarchies [1]:
- Random access to menu items (Key shortcuts).
- Pointer setting strategies for popup menus provide for a faster access to a certain menu item, often the most recently or most frequently used.
- Offering user support in selecting valid and appropriate items (Grey-shading of disabled menu-items).
- Action prompting according to previously selected objects (object-specific menus or dynamically exchanged control panels).
- Reorganization of menus according to user-specific usage patterns [2].
- Automating iterative patterns in interaction. For example Eager which is a Programming-By-Example system that anticipates which action the user is going to perform next [3].

We propose an agent-based architecture to tackle the problem described above. Adaptive agents, or adaptively communicating concurrent modules making automatic decisions on behalf of the user, are viewed by many as a way of decreasing the amount of human-computer interaction necessary for the information management task [4].

Beale and Wood [5] identify the general properties of an agent and consider what makes a task "Agent-worthy":

- Adapting --- the task requires a degree of adaptability, so that the software has to refine its skills to learn new or better ways of accomplishing things. This includes learning to avoid failure and accommodating user preferences. Using AI techniques, adaptive agents are able to judge their results, then modify their behavior (and thus their internal structure) to improve their perceived fitness.
- Researching --- the task is poorly defined, so the software has to explore a number of different options based on its current expertise.
- Demonstrating --- the task involves passing on skills that the software uses. This encompasses the software teaching users how to do things, and also provides explanations of what the agent is up to. It also allows one agent to teach another.
- Guiding --- the task requires some degree of cooperation between the agent and the user. The software could offer constructive criticism to the user as they work, or could assist them in working though a particular set of options.
- Autonomy --- the task itself requires regular or constant attention, but not necessarily any user input, thus making delegation useful. An example of this is the monitoring of the System State with certain events triggering specific actions.
- Asynchrony --- the task has a significant delay between its initiation and completion. These delays can be due to slow processing or to network delays, or to temporary unavailability of the required service.

We can therefore conclude that human-computer interaction is a suitable candidate for agent-based solutions.

Some examples of agent-based interaction are sited below:
- Agents can be used to allow the customized presentation of information [8]. These agents preprocess the data and display it in a way that can be unique for each individual user.
- Agents that act as tutors or guides, supplementing user's knowledge with their own [9]. These assist the current task by providing alternative views and additional relevant information [7].
- Agents could be used for the adaptive search and retrieval of information [10].

A predominant approach to the use of Agents in interaction has been to concentrate a large bulk of the interaction responsibilities on a single agent, thus reverting to a centralized architecture. Nevertheless many real world problems are best modeled using a set of cooperating intelligent systems (agents) [6][11]:
- Our society consists of many interacting entities and so if we are interested in modeling some aspects of our society, our model needs to be structured.
- Data often originates at different physical locations; therefore centralized solutions are often inapplicable or inconvenient.
- Using a number of small simple adaptive agents instead of one large complicated one may simplify the process of solving a complicated problem [12]. In other words agents collectively exhibit emergent behavior, where the behavior of the agent population as a whole is greater than the sum of its parts.

In this paper we will propose an adaptive multi-agent based interaction method. First we will give a description of the Adaptive Agent Oriented Software Architecture (AAOSA) [13]. Then we will describe an implementation of AAOSA for interactive systems using a simple example. A simple distributed learning algorithm is presented next, after which an overall evaluation of the proposed system will be given. Finally, some suggestions will be made for future work in this area.

**2. Adaptive Agent Oriented Software Architecture**

The classic view taken with respect to Agent Oriented Systems is to consider each agent an autonomous individual the internals of which are not known and that conforms to a certain standard of communications and/or social laws with regard to other agents [15]. Architectures viewing agents as such have had to introduce special purpose agents (e.g., broker agents, planner agents, interface agents…) to shape the structure into a unified entity desirable to the user [14]. The intelligent behavior of these key agents, with all their complexities, would be vital to the performance of the whole system.

On the other hand, methodologies dealing with the internal design of agents tend to view them primarily as intelligent, decision-making beings. In these methodologies, techniques in Artificial Intelligence, Natural Language Processing, machine learning, and adaptive behavior seem to overshadow the agent's architecture, in many cases undermining the main purpose of the agent [16][17].

Instead of using assisted coordination, in which agents rely on special system programs (facilitators) to achieve coordination [19], in the AAOSA approach new agents supply other agents with information about their capabilities and needs. To have a working system from the beginning, the designers preprogram this information at startup. This approach is more efficient because it decreases the amount of communication that must take place, and does not rely on the existence, capabilities, or biases of any other program [18].

Adaptive agents are adaptively communicating, concurrent modules. The modules therefore consist of three main parts: A communications unit, a reward unit, and a specialized processing unit. The first two units we will call the white box and the third the black box parts of an agent (Figure 1). The white box part of an agent is common in all AAOSA agents although some functions may be left unused in certain cases. From our point of view, the black box is simply unknown and completely left to the designer.
The main responsibilities of each unit follow:

*The communications unit:* This unit facilitates the communicative functions of the agent and has the following sub-systems:

- *Input of received communication items:* These items may be in a standard agent communication language such as KQML.
- *Interpreting the input:* Decides whether the process unit will need or be able to process certain input, or whether it should be forwarded to another agent (or agents). Note that it is possible to send one request to more than one agent, thus creating competition among agents.
- *Interpretation Policy:* (e.g., a table) Determines what should be done to the input. This policy could be improved with respect to the feedback received for each interpretation from the rewards unit. Some preset policy is always desirable to make the system functional from the beginning. In the case of a system reset, the agent will revert to the basic hard-coded startup information. The interpretation policy is therefore comprised of a preset knowledge base, and a number of learned knowledge bases acquired on a per-user basis. A *learning* module will be responsible for conflict resolutions in knowledge-base entries with regard to feedback received on the process of past requests. Past requests and what was done with them are also stored in anticipation of their feedback.
- *Address-Book:* keeps an address list of other agents known to be useful to this agent, or to agents known to be able to process input that can not be processed by this agent. Requests to other agents may occur when:
  - The agent has received a request it does not know how to handle,
  - The agent has processed a request and a number of new requests have been generated as a result.
  
  This implies that every agent have an address and there be a special name server unit present in every system to provide agents with their unique addresses (so that new agents can be introduced to the system at run time). This address list should be dynamic, and therefore adaptive. It may be limited; it may also contain information on agents that normally send their requests to this agent. In many cases the Address-book could be taken as an extension of the Interpretation Policy and therefore implemented as a single module.
- *Output:* Responsible for sending requests or outputs to appropriate agents, using the Address-book. A confidence factor could be added to the output based on the interpretations made to resolve the input request or to redirect it. We shall see later in the paper that this could be used when choosing from suggestions made by competing agents by output agents.

*The rewards unit:* Two kinds of rewards are processed by this module: outgoing and incoming. An agent is responsible for distributing and propagating rewards being fed back to it[*]. This unit will determine what portion of the incoming reward it deserves and how much should be propagated to requesting agents. The interpreter will update its interpretation policy using this feedback. The rewards will also serve as feedback to the Address-book unit, helping it adapt to the needs and specifications of other agents. The process unit could also make use of this feedback.

The rewards may not be the direct quantification of user states and in most cases will be interpretations of user actions made by an agent responsible for that. We will further clarify this point later in the paper.

*The process unit:* This unit is considered a black box by our methodology. The designer can use whatever method it deems more suitable to implement the processes unique to the requirements of this agent. The only constraints being that the process unit is limited to the facilities provided by the communications unit for its communications with other agents. The process unit also may use the rewards unit to adapt its behavior with regard to the system. Note that each agent may well have interactions outside of the agent community. Agents responsible for user I/O are an example of such interactions. These agents generally generate requests or initiate reward propagation in the community, or simply output results.

The white box module can easily be added to each program module as a *transducer*. According to definition [18] the transducer mediates between the existing program (the process unit) and other agents.

---

[*] A special purpose agent is responsible for the interpretation of user input as feedback to individual user requests. This agent will then initiate the reward propagation process.

The advantage of using a transducer is that it requires no knowledge of the program other than its communication behavior.

We mentioned the process unit as being able to conduct non-agent I/O. It is easy to consider I/O recipients (e.g. files or humans) as agents and make the program redirect its non-agent I/O through its transducer. Other approaches to agentification (wrapper and rewriting) are discussed in [18].

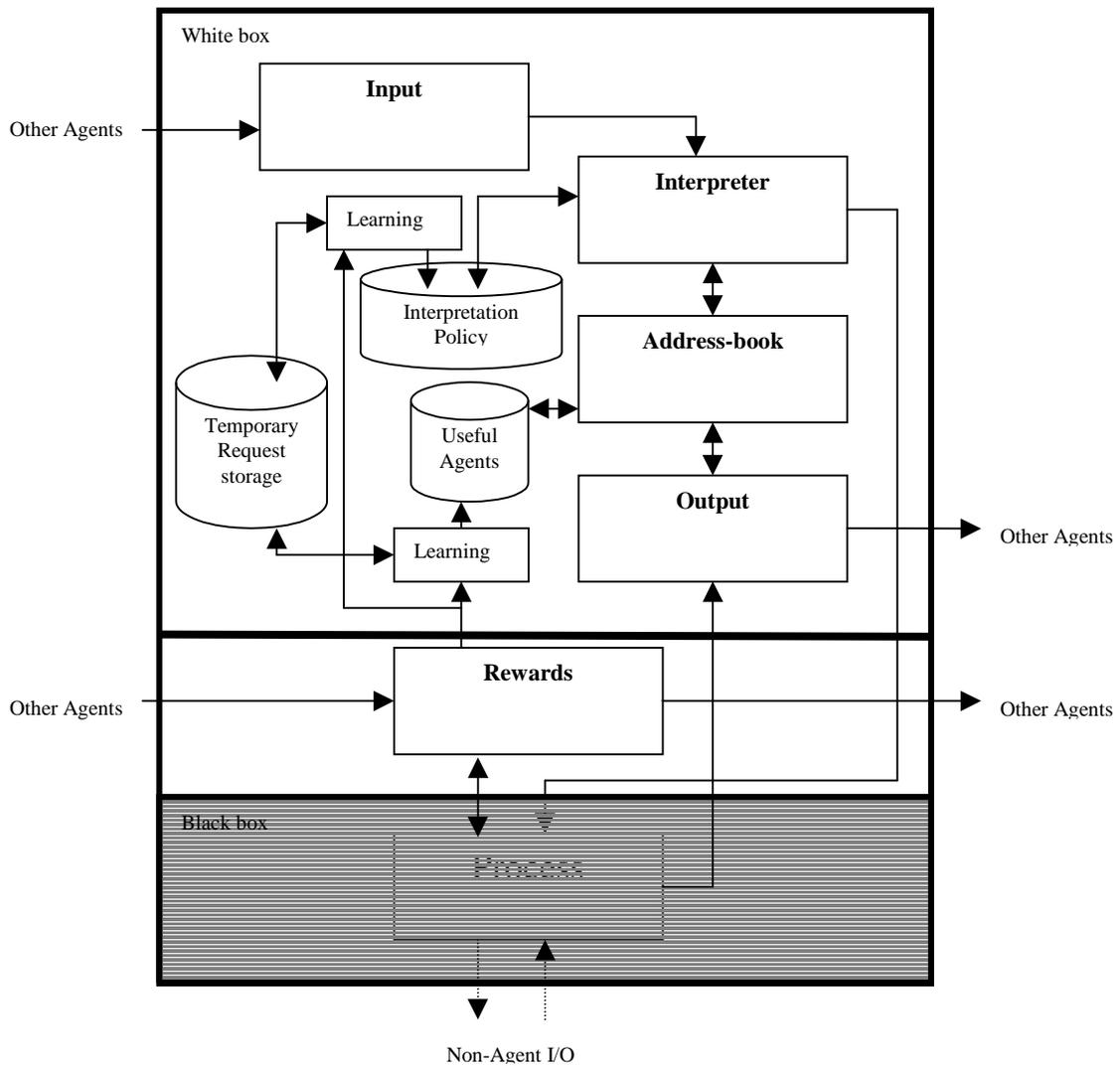

**Fig. 1.** Each agent is comprised of a black box section (specialties) and a white box section (communications).

## 3. Design issues

The designer or design team should see the software as a whole and therefore the responsibility of the software's proper functionality lies fully with the designers. Only their design can guarantee the soundness and correctness of the system, although the dynamism, flexibility and adaptive nature of AAOSA will help to a great extent in facilitating the processes of designing and debugging a software application.

The software as a whole should be thought of as a society, striving to accomplish a set of requests. The input requests are therefore propagated, processed by agent modules that may in turn create requests to other agents. Again, it is up to the designers to break down the system, as they feel suitable. Hierarchies of agents are possible and agents can be designed to be responsible for the minutest processes in the system. It is advisable that each agent be kept simple in its responsibilities and be limited in the decisions it needs to make to enhance its learning abilities. The overhead of the required units (the white box) should be taken into consideration.

Agents can be replaced at run-time with other more complete agents. The replacement can even be a hierarchy or network of new agents breaking down the responsibilities of their predecessor. This feature provides for the incremental design and evaluation of software.

We recommend each agent's responsibilities and actions to be clearly defined at design time. As stated in the previous section, many aspects of the white box units should also be preset for each agent according to its definition. To have a working system from the beginning, it is also necessary to define the preliminary communication links between the agents at design time. It should be noted that these communications might change through time, for instance in the case of the introduction of newer agents. Thus, the most important phase in the design of software with this methodology will be to determine the participating agents and their capabilities; even though the precise details of how they will eventually communicate with each other may not be known at design time.

### 3.1 Special purpose agents
Some special purpose agents may be used depending on the application, for example, agents directly involved with initiating requests to the system, or an agent which interprets the actions of the user as different levels of reward.

*Input Agents*
Inputs to the system may be from more than one source. In such systems, one, or a network of special purpose input agents should be considered, which:

- Unify inputs from different sources into one request, and/or
- Determine commands that all fall into a single request set.

For example if the user's Natural Language (NL) input is: "Information on this" and the mouse is then pointed at an item on the display, these two inputs should be unified as a single request.

Interactive input would also require the input agents to determine the end of an input stream. For instance in NL input a phrase (e.g., Do! or Please!) or a relatively long pause, could determine the end of the stream. A different policy here would be to interpret the input in real-time and redirect it to the appropriate agent. As seen in figure 2, agents can redirect the input back to the input agents once this data is no longer relevant to the responding agent.

*Feedback Agents*
Any adaptive system needs a reward feedback that tells it how far from the optimum its responses have been. This reward could be explicitly input to the system, or implicitly judged from input responses by the system itself. In the case of implicit rewarding, an agent could be responsible for the interpretation of the input behavior and translating it into rewards. The criteria that could be used depend on the system. For instance in an interactive software application a repeat of a command, remarks indicating satisfaction or dissatisfaction, user pause between requests or other such input could be translated into rewards to different output.

The feedback agents could also be more than one depending on the different judgement criteria and a hyper-structure [12] or hierarchy might be designed to create the rewards.

A name server unit is also required to make possible the dynamic introduction of new agents to the system. Each new agent will have to obtain its name (or address) from this name server so that conflicts do not occur and so agents can be referred to throughout the system. Input requests (commands) to the system should be tagged with the user-id that has issued them because interpretation is done on a per-user basis. Reward fed back to the system should also incorporate the request-id to which the reward belongs.

**3.2 Communication Language**

Communication may supply the agent with valuable information and thus avoid "re-discovering the wheel" [11]. In Agent-Based Software Engineering [18], Agents receive and reply to requests for services and information using a declarative knowledge representation language KIF (Knowledge Interchange Format), a communication language KQML (Knowledge Query and Manipulation Language) and a library of formal ontologies defining the vocabulary of various domains. We will give a basic description of the KQML extension we have defined in an example application in the following section.

**4. Modeling an interaction system using AAOSA**: A multimodal map [20]

Multiple input modalities may be combined to produce more natural user interfaces. To illustrate this technique Cheyer and Julia [20] present a prototype map-based application for a travel-planning domain. The application is distinguished by a synergistic combination of handwriting, gesture and speech modalities; access to existing data sources including the World Wide Web; and a mobile handheld interface. To implement the described application, a distributed network of heterogeneous software agents was augmented by appropriate functionality for developing synergistic multimodal applications.

We will consider a simplified subset of this example to show the differences of the two approaches. A map of an area is presented to the user and she is expected to give view port requests (e.g., shifting the map or magnification), or request information on different locations on the map. For example, a user drawing an arrow on the map may want the map to shift to one side. On the other hand the same arrow followed by a natural language request such as: "Tell me about this hotel." May have to be interpreted differently.

Cheyer and Julia [20] use the Open Agent Architecture (OAA) [21] as a basis for their design. In this approach, based on a "federation architecture" [22], the software is comprised of a hierarchy of facilitators and agents. The facilitators are responsible for the coordination of the agents under them so that any agent wanting to communicate with any other agent in the system must go through a hierarchy of facilitators (starting from the one directly responsible for it). Each agent, upon introduction to the system, provides the facilitator above it with information on its capabilities (Figure 2). No explicit provision is given for learning.

An example design based on AAOSA is shown in figure 3. It must be noted here that the design shown here is not rigid and communication paths may change through time with the agents adapting to different input requests.

The text and pointer input agents determine the end of their respective inputs and pass them on to the input regulator. This agent in turn determines whether these requests are related or not. It then passes them down to the agent it considers more relevant. The output agents simply actuate suggestions made by the shifting, magnification, hotels, restaurants and general information agents. Note that combinations of these output suggestions could also be chosen for actuation. The feedback agent provides the system with rewards interpreted from user input.

Some of the differences in the two designs are given below:

- The AAOSA design is much more distributed and modular by nature and many of the processes concentrated in the facilitator agents in figure 2 are partitioned and simplified in figure 3.
- AAOSA is more of a network or hyper-structure [12] of process modules as opposed to the hierarchical tree like architecture in the OAA design.
- AI behavior such as input interpretation (e.g., natural language processing) and machine learning are incorporated on the architecture and distributed over the multi-agent structure rather than introduced as single new agents (as is the case with the natural language macro agent in figure 2).

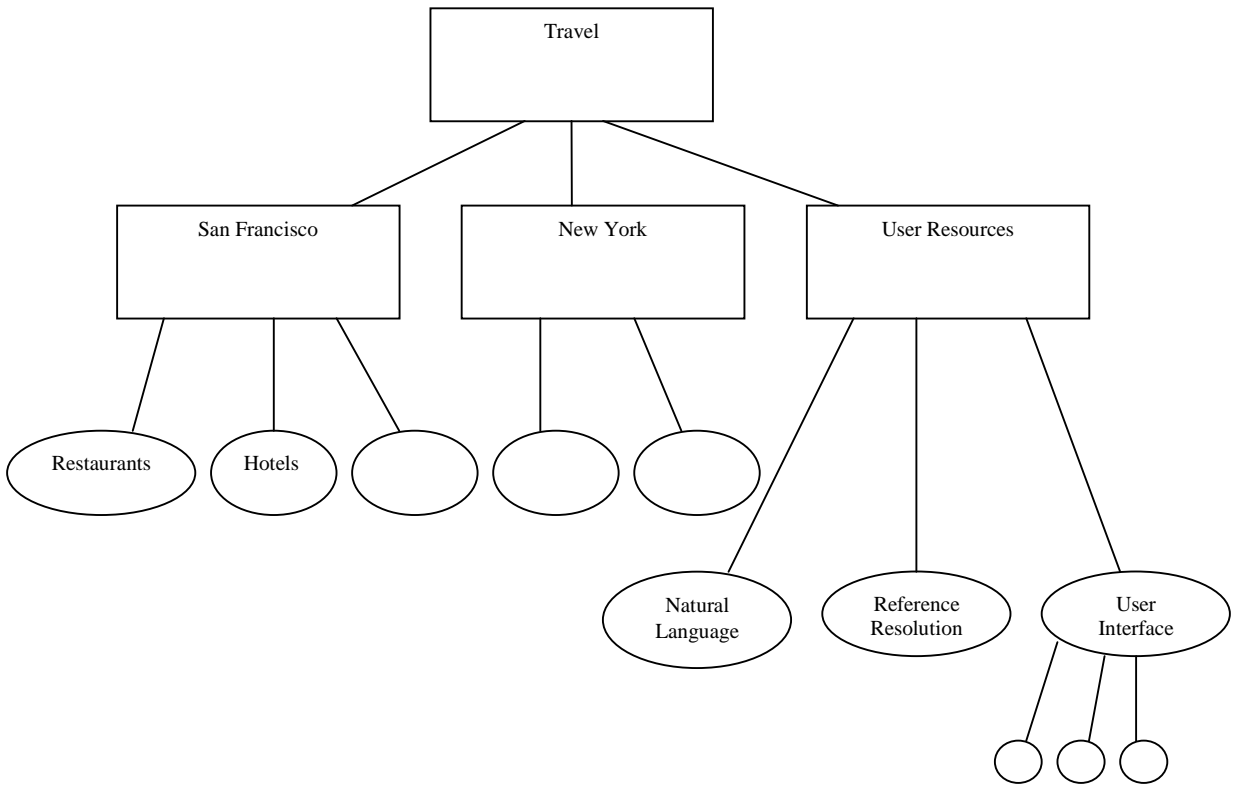

**Fig. 2.** A structural view of the multimodal map example as designed using OAA in [20]. Boxes represent Facilitators, ellipses represent macro agents and circles stand for modality agents.

It must be stressed that AAOSA like architectures could be achieved with an OAA if we take each OAA facilitator and its macro agents as one agent and add learning capabilities to each facilitator. Another point worth mentioning is that agents in OAA are usually pre-programmed applications linked together through facilitators. The designers have a lesser say over the software architecture as a whole because they are forced to use what has already been designed, possibly without the new higher-level framework in mind.

In designing an interactive system using AAOSA we have assigned an agent to each individual function of the system (e.g., Magnification, Shifting, Hotel information, Restaurant Information, General Information). The functionality of these agents is implemented in the black box of each agent. These agents are also responsible for maintaining a representation of their respective domain. For instance the Magnification Agent maintains a variable representing the current degree of magnification.

The structure of agents that lead to these leaf agents represent the designer's view of the system hierarchy. These agents usually have a much simpler black box. Their role is mainly to direct requests to the appropriate agents and learn or resolve contradictions that may occur at their juncture.

**AAOSA Interaction System's communication performatives**

In this section we will describe the common performatives used by the agents to communicate. These performatives are all general and therefore pre-implemented in the white box modules of these AAOSA agents.

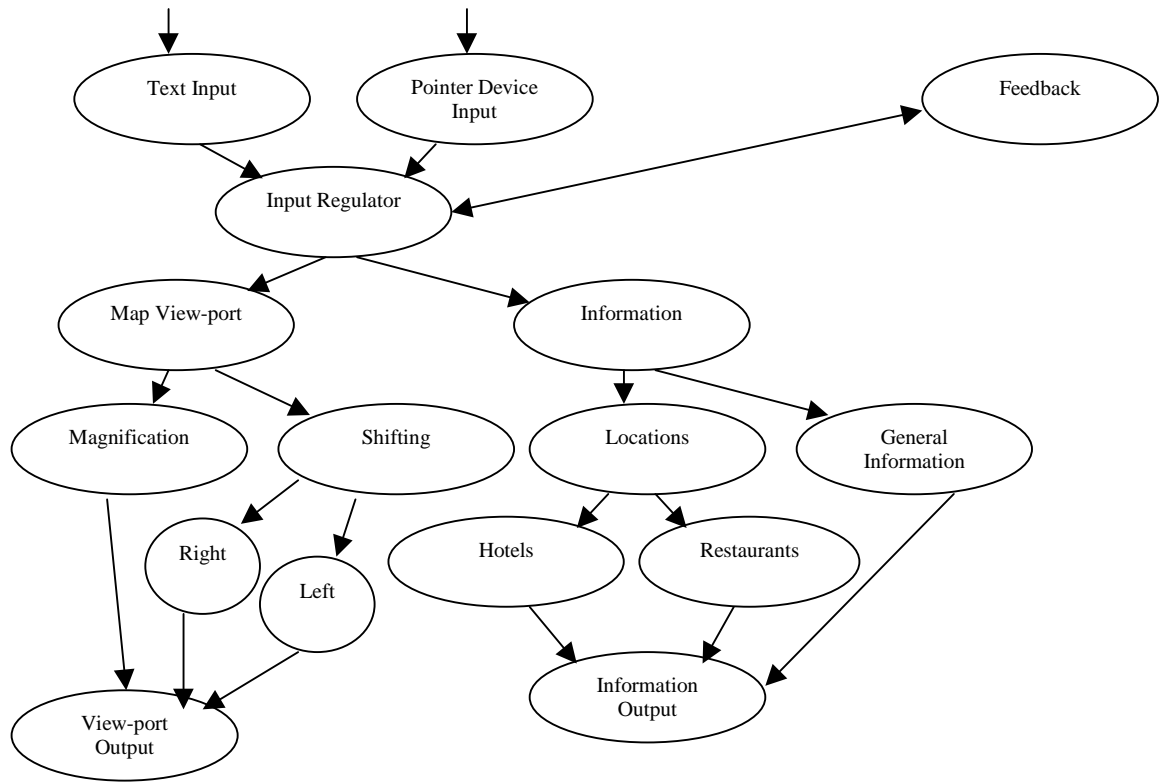

**Fig. 3.** The multimodal map example as designed based on AAOSA.

Let us follow a sample run of the multimodal map example:

REGISTER
Agents need to register themselves with each other to be able to send messages to one another. Unlike similar systems, in AAOSA all agents need not be aware of all other agents and registration may be much more distributed and localized. For instance the information agent need only to register with the Location, General information, and Input regulator agents. Each agent, upon receiving a Register message, adds the registering agent's name and address in its address book. Registration may take place at run time.

ADVERTISE and UN-ADVERTISE
Agents advertise their responsibilities in order to draw requests from other agents. When an agent receives an advertise message it updates its interpretation policy so as to redirect certain requests to the advertising agent. This information is removed from the interpretation policy once an un-advertise message is received. An advertise message specifies a community to which the advertising agent belongs. If the agent receiving this message does not recognize such a community in its interpretation policy, it may add it as a new community. In the multimodal map example shifting agent advertises "shifting" to Map View-port agent which in turn creates a new community by that name in its interpretation policy. This allows for more than one agent being members of an interpretation community of another agent.

THIS-IS-YOURS
When an agent is successful in interpreting a request, it must be passed over to an agent from the interpreted community. The performative under which this request is forwarded is called THIS-IS-YOURS. The receiving agent knows that if it can not interpret this request, then the point of contradiction is itself. For example let us say the Input regulator agent receives "Map to the right" from the text-input agent. If the interpretation policy for routing requests to the Map view-port community is simply the presence of the word "map" in the requests, this request is sent to an agent in that community. In our example we only have one agent per community so a THIS-IS-YOURS message with the request as its content will be sent to the Map view-port agent.

IS-THIS-YOURS?
IT-IS-MINE
NOT-MINE

For an Agent-based interactive system, adaptation occurs to accommodate user preferences, whilst the conditions that the task must satisfy are poorly defined thus necessitating a search to find a suitable solution [5]. As we mentioned earlier, the interpretation of input is distributed over the structure of agents so there may well be situations in which an agent can not directly interpret a request and will have to query the communities it knows and wait for their response. Each agent itself may query agents belonging to its communities. In our example, using the same simplistic interpretation method (i.e., presence or absence of key words), when the Map view-port agent receives "Map to the right" it can not interpret it and so it will query the communities it knows by sending them IS-THIS-YOURS? messages. It will then wait until it has received responses from all agents it has queried. The Magnification agent, not having any communities, will respond with NOT-MINE. The Shifting agent, having interpreted the request successfully (because of the presence of "right" in the request string), responds with IT-IS-MINE. At this point the Map view-port agent sends the query down to the Shifting agent using the THIS-IS-YOURS performative.

**5. Contradictions**

An important issue is how to combine the opinions of different agents. Generally certain confidence factors are associated with each decision and then some method is used to generate the final decision on the basis of the individual decisions. If an agent is capable of dealing with a subset of given problems, and if this agent can be considered "sufficiently reliable", we do not need to worry about redundancy at all. In this case the answer of that one agent is sufficient [11].

Contradictions occur in the following cases:
- When an agent that has received a THIS-IS-YOURS message can not interpret it and all of its communities upon being queried respond with NOT-MINE messages.
- When an agent that has received a THIS-IS-YOURS message can not interpret it and more than one of its communities upon being queried responds with an IT-IS-MINE message.
- When the user expresses dissatisfaction with a response from the system.

In our implementation of the interpretation system we have used a combination of the following methods to resolve such contradictions:

- Using priorities or weights for agents of different levels in the hyper-structure. For instance in a contradiction that has occurred in the Input regulator agent, the IT-IS-MINE that originates in the Locations agent will have a higher priority than the one originating in the Left agent.
- Querying the user directly and adjusting the interpretation policy accordingly (i.e., learning). Note that in this case the interaction is limited to the point of contradiction.

Contradiction plays an important role in learning and pin pointing the agent which is responsible for a contradiction and resolving it there insures the correct distribution of responsibilities in the hyper-structure.

**6. Distributed Learning**

The problems of multi-agent learning have been largely ignored. Designing agents that would learn about anything in the world goes against the basic philosophy of distributed AI. This issue has not really been paid attention to. This may be the reason why some systems are ill behaved (the more they learn the slower they perform) [11]. Allowing the agents to adapt, refine and improve, automatically or under user control, we can create a holistic system in which the whole is significantly more than the sum of its parts [5].

The combination of machine learning and multi-agent systems can have benefits for both. Multi-agent systems having learning capabilities will reduce cost, time, and resources and increase quality in a number of forms [11]:
- Ease of programming
- Ease of maintenance
- Widened scope of application
- Efficiency
- Coordination of activity

On the other hand machine learning in a multi-agent set-up becomes faster and more robust [11].

Adaptability in AAOSA materializes in the following forms:
- The ability of the system to accept new agents at run time,
- The ability of each agent to adapt its behavior according to the feedback it receives (i.e., learning).

A sample of the run of the program and the contradiction resolution process is described in Figure 4. The learning algorithm in the simplest form could be a memorization of the user response. This method lacks generalization and is context independent. In other words the interpretation can not be modified to include the previous state of the system.

1) User inputs "move it closer"
   a) "move it closer" is sent to the Input regulator Agent with a "Is-This-Yours" performative.
   b) The Input regulator agent in turn sends "move it closer" down to the Information community and the Map View-port community with an "Is-This-Yours?" performative.
   c) Agents in the communities beneath the Information all respond with "Not-Mine" and therefore the Information agent also sends a "Not-Mine" message to the Input regulator agent.
   d) The Magnification agent and the Shifting agent both claim "move it closer" to be theirs by sending "It-Is-Mine" messages to the Map View-port agent.
   e) The Map View-port agent announces a contradiction by sending a "Maybe-Mine" message to the Input regulator agent.
   f) The Input regulator agent sends a message to the Map View-port agent asking it to "resolve" its contradiction because it has not had any other positive responses from agents with a higher priority.
   g) The Map View-port agent interacts with the user to resolve the contradiction:
2) System Asks User: "Do you mean magnification or shifting?"
3) User responds with "Magnification"
   a) The Map View-port agent learns that "move it closer" belongs to the Magnification agent and resolves the contradiction.
   b) The Map View-port agent sends "move it closer" to the Magnification agent with "This-Is-Yours".
   c) The Magnification agent interprets "move it closer" and sends a "This-Is-Yours" performative with "zoom in" as the content to the View-port output agent.
4) View-port is magnified

**Fig. 4.** A sample run of the Multi-modal map example.

**7. Evaluation**

Viewing software as a hyper-structure of Agents (i.e., intelligent beings) will result in designs that are much different in structure and modularization. Some of the benefits of this approach are noted here.

- Flexibility: There is no rigid predetermination of valid input requests.
- Parallelism: The independent nature of the agents creates a potentially parallel design approach.
- Multi platform execution: Agents can run and communicate over networks of computers (on the Internet for instance).
- Runtime addition of new agents and thus incremental development of software.
- Software additions by different designers: Different designers can introduce different agents to compete in the software, making this design methodology attractive for commercial applications.
- Reusability of agents.
- Incremental design and evaluation.

- Learning and Intelligence: The distributed nature of learning introduced in this paper suggests a powerful adaptive software design that potentially breaks down an application to a hyper-structure of simple learning modules [12].

## 8. Future Work

Although the use of AAOSA architecture in the area of interactive systems seems to be promising, this methodology is still at its infancy and therefore should be tested for its scope of applicability. Two main areas in AAOSA have the potential for greater improvement:

- Each agent's learning algorithm can be improved immensely. Even so, in our simple implementation the fact that the learning was distributed resulted in an effective system. Improvements could be made to provide for:
  - Generalization,
  - Context sensitivity,
  - Acquisition of undefined concepts [23].
- Using AAOSA we were able to successfully localize the responsibilities of a society of agents collaborating in an interactive application. This localization is vital if we want to use distributed learning effectively because it helps limit the complexity each agent has to deal with to the scope pre-designed for that agent. The problem here is that the success of this localization depends on the designers. One improvement to the current system is to have the initial design evolve and optimize at run time. This optimization could be done using evaluational feedback from different sources that give an estimate for the complexity the different agents should deal with. The estimate can be made using:
  - Explicitly from the user,
  - Implicitly from the user (e.g., The feedback agent can interpret user actions),
  - From the rate of contradictions occurring in different agents,
  - Time or resources.

  According to the feedback new agents could be added to the system once the points of highest complexity are found. The ultimate would be for the agents to be able to split their responsibilities and knowledge (like a single cell organism dividing). On the other hand using evolutionary methods we can remove unwanted and redundant agents passing their responsibilities to other agents.